\documentclass[onecolumn,showpacs,preprintnumbers]{revtex4}
\usepackage{amsthm, amscd, amsfonts, amssymb, graphicx}
\usepackage{dcolumn}
\usepackage{bm}
\usepackage[english]{babel}

\begin{document}

\title{
Impurities with retarded interaction with quasiparticles in a
s-wave superconductor.}
\author{K.V. Grigorishin}
\email{gkonst@ukr.net} \affiliation{Boholyubov Institute for
Theoretical Physics of the National Academy of Sciences of
Ukraine, 14-b Metrolohichna str. Kiev-03680, Ukraine.}
\date{\today}

\begin{abstract}
A perturbation theory and a diagram technique for a disordered
metal are proposed when scattering of quasiparticles by
nonmagnetic impurities is caused with a retarded interaction. The
perturbation theory generalizes a case of the elastic scattering
in a disordered metal. Eliashberg equations for s-wave
superconductivity are generalized for such a disordered
superconductor. Anderson's theorem is found to be violated in the
sense that embedding of the impurities into a s-wave
superconductor increases its critical temperature. We showed the
amplification of superconducting properties is a result of
nonelastic effects in a scattering by the impurities.

\end{abstract}

\keywords{disordered s-wave superconductor, impurity, Anderson's
theorem, retarded interaction, Eliashberg equations,
superconducting transition temperature}

\pacs{74.62.En,72.80.Ng} \maketitle

\section{Introduction}\label{intr}

As is well known a s-wave superconducting state is stable regard
to embedding of nonmagnetic impurities. In this case an ordinary
potential scattering acts on both electrons of a Cooper pair
equally, therefore the pair survives. Mathematically this is
expressed in the fact that a gap $\Delta$ and an energetic
parameter $\varepsilon$ are renormalized the same manner:
$\frac{\widetilde{\Delta}}{\widetilde{\varepsilon}}=\frac{\Delta}{\varepsilon}$,
where $\widetilde{\Delta},\widetilde{\varepsilon}$ are the
renormalized values by an impurity scattering. As a consequence a
critical temperature of a superconductor does not change. This
statement is Anderson's theorem \citep{ander,sad,sad1}. However, a
strong suppression of superconductivity takes place near
Anderson's transition metal-insulator, that is when
$\frac{1}{k_{F}l}\gtrsim 1$, where $l$ is a free length and
$k_{F}$ is Fermi momentum. Although in a state of Anderson's
isolator a superconductive response of the system can be remain
\citep{sad,sad1,bulaev}. The magnetic impurities differently acts
on components of Cooper pair, with the result that its decay takes
place. Superconducting state is unstable regard to embedding of
magnetic impurities - the critical temperature decreases that is
accompanied by effect of gapless superconductivity
\citep{sad,degenn}. For $d$-wave superconductors the nonmagnetic
impurities destroy superconductivity like magnetic impurities
\citep{bork,fehren,radt,posazh,loktev,pogor}. The reduce of the
critical temperature is a mathematical consequence of an
inequality
$\frac{\widetilde{\Delta}}{\widetilde{\varepsilon}}<\frac{\Delta}{\varepsilon}$,
that is the gap and the energetic parameter are renormalized in
different ways. It should be noticed if electrons are paired with
nonretarded interaction (as in BCS theory, negative U Hubbard
model) then the superconductive order parameter strongly
suppressed with an increase of disorder \citep{fay,mor}. This
means we must use approaches which take into account the fact that
quasiparticles are paired with retarded interaction (for example
with electron-phonon interaction, electron-magnon interaction
etc.)

In a work \citep{grig} a case was considered when when scattering
of quasiparticles by nonmagnetic impurities is caused with a
retarded interaction. The retarded interaction occurs because the
impurities have an internal structure and make transitions between
their states under the action of metal's quasiparticles. In the
proposed model the principal possibility of increasing of the
critical temperature due to the retarded interaction between
quasiparticles and impurities has been shown. However a formal
retarded form of the electron-impurity interaction was proposed
only and an electron-impurity coupling constant has not been
calculated. Thus the theory does not enables us to calculate the
critical temperature since an internal structure of the impurities
and its interaction with quasiparticles is unknown. In addition an
impurity was supposed as a two-level system with an eigenfrequency
$\omega_{0}$ and the simplest type of the diagrams was considered
only.

This paper is aimed to generalize disordered metal's theory when
interaction of quasiparticles with nonmagnetic impurities is
retarded. In Section \ref{normal} we develop a diagram technique
for a disordered metal in normal state. Main types of diagrams are
determined and contributions of the each type are estimated. A
perturbation theory is made using an adiabaticity parameter and an
method of an uncoupling of correlations. In Section \ref{super} we
develop a diagram technique for a disordered metal in s-wave
superconducting state. Eliashberg equations are generalized to a
case when the superconductor contains impurities of a considered
type. Based on these equations we show the retarded interaction of
quasiparticles with impurities violates Anderson's theorem in the
direction of increasing of the superconducting transition
temperature.

\section{Normal state.}\label{normal}

\begin{figure}[h]
\includegraphics[width=8.0cm]{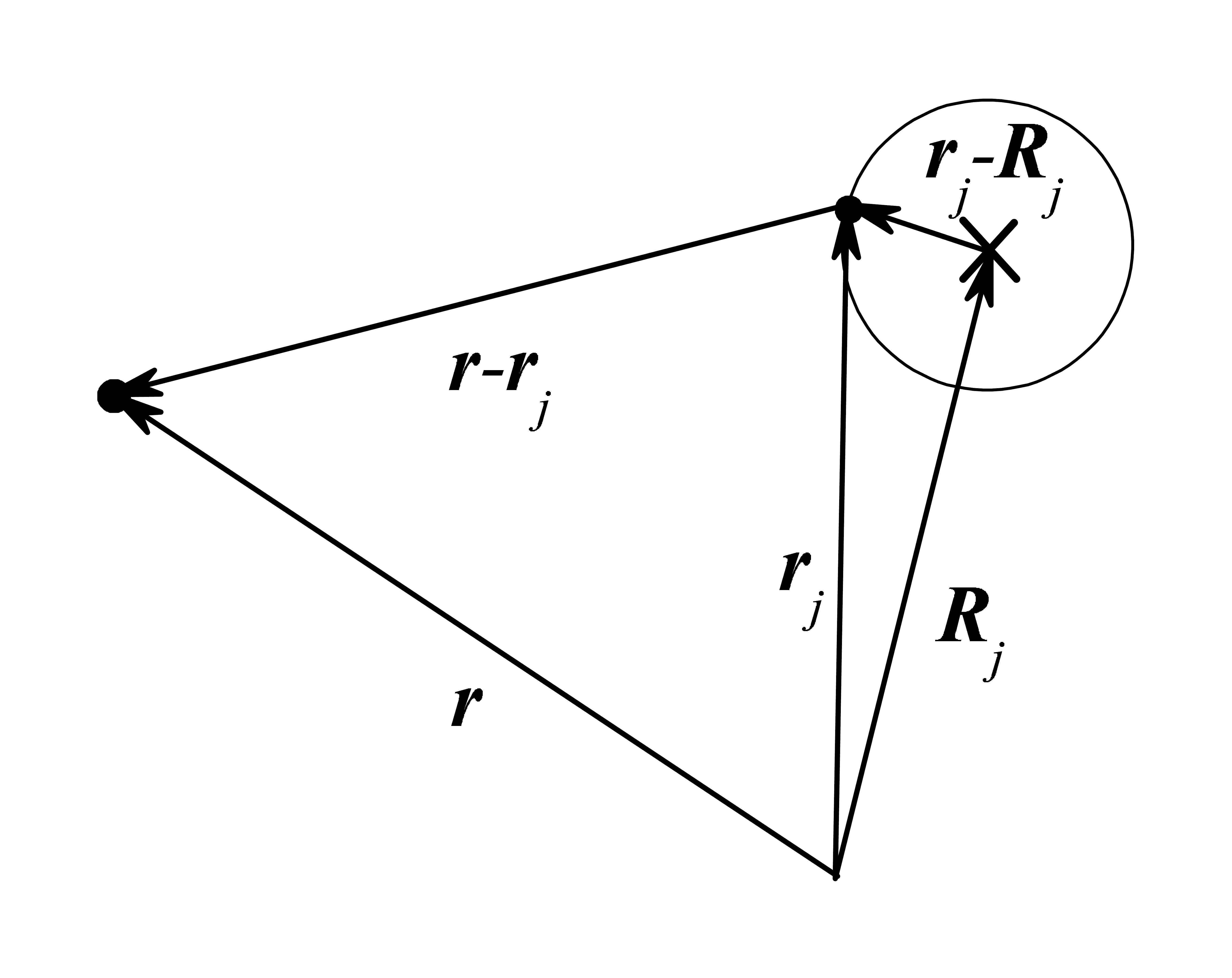}
\caption{A mutual disposal of a quasiparticle with a radius-vector
$r$ and a $j$th impurity with a radius-vector $\textbf{R}_{j}$.
Interaction between the quasiparticle and the impurity is
$U(\textbf{r}-\textbf{r}_{j})$, where $\textbf{r}_{j}$ is a
radius-vector of a state configuration
$\phi\left(\textbf{r}_{j}-\textbf{R}_{j}\right)$ of the $j$th
impurity.} \label{Fig1}
\end{figure}

Let an electron moves in a field created by $N$ scatterers
(impurities) which are placed in points $\textbf{R}_{j}$ by a
random manner with concentration $\rho=\frac{N}{V}$. Each impurity
can be in states
$\phi_{A}\left(\textbf{r}_{j}-\textbf{R}_{j}\right),\phi_{B}\left(\textbf{r}_{j}-\textbf{R}_{j}\right),
\phi_{C}\left(\textbf{r}_{j}-\textbf{R}_{j}\right),\ldots$ with
energies $E_{A}, E_{B}, E_{C},\ldots$ accordingly. Here
$\textbf{r}_{j}$ is a radius-vector of a state configuration
$\phi$ of $j$th impurity (Fig.\ref{Fig1}).
Metal's quasiparticles (conduction electrons) are described with
wave functions
$\psi_{\textbf{k}}(\textbf{r})=\frac{1}{\sqrt{V}}e^{i\textbf{kr}}$
and fill all states up to Fermi momentum $k_{F}$. Interaction of
the electrons with the impurities is described with potential
$U(\textbf{r}-\textbf{r}_{j})$. As a result of scattering the
electrons go into states $\psi_{\textbf{k}'}(\textbf{r})$, and the
impurities go from some state $\phi_{A}$ into one of states
$\phi_{A,B,C,\ldots}$. Main approximation lies in the fact that
oscillations of an impurity are local, that is
$E_{B}-E_{A}\equiv\omega_{BA}\neq\omega(q)$, where $\textbf{q}$ is
a wave vector. On the contrary a phonon frequency is a function of
a wave vector $\Omega=\Omega(q)\leq\Omega_{D}$ (here $\Omega_{D}$
is Debye frequency). The impurity's oscillations do not interact
with phonons of a metal. Then Hamiltonian of the system can be
written as follows:
\begin{eqnarray}\label{1.1}
\widehat{H} &=&
\widehat{H}_{0}+\sum_{j}\widehat{H}_{0}^{j}+\sum_{j}\sum_{A,B}\sum_{\textbf{k},\textbf{k}'}\int\int\phi_{B}^{+}
\left(\textbf{r}_{j}-\textbf{R}_{j}\right)\psi_{\textbf{k}'}^{+}(\textbf{r})
U(\textbf{r}-\textbf{r}_{j})\phi_{A}\left(\textbf{r}_{j}-\textbf{R}_{j}\right)\psi_{\textbf{k}}(\textbf{r})d\textbf{r}d\textbf{r}_{j}
c^{+}_{B}c^{+}_{\textbf{k}'}c_{A}c_{\textbf{k}},
\end{eqnarray}
where $\widehat{H}_{0}$ is Hamiltonian of a homogeneous medium
without impurities, $\widehat{H}_{0}^{j}$ is Hamiltonian of a
$j$th impurity:
$H_{0}^{j}\phi_{A}\equiv\left(\frac{\widehat{p}^{2}}{2M}+V(\textbf{r})\right)\phi_{A}=E_{A}\phi_{A}$,
$c_{A}$ and, $c^{+}_{B}$ are creation and annihilation operators
of an impurity in states $\phi_{A,B,C,\ldots}$, $c_{\textbf{k}}$
and $c^{+}_{\textbf{k}'}$ are creation and annihilation operators
of an electron in states $|\textbf{k}\rangle$ and
$|\textbf{k}'\rangle$. The third term describes interaction of
electrons with impurities. A free propagator of electrons is:
\begin{eqnarray}\label{1.2}
G_{0}(\textbf{k},\varepsilon)=\frac{1}{\varepsilon-\xi(k)+i\delta\texttt{sign}\xi}
\end{eqnarray}
where $\xi(k)=\frac{k^{2}}{2m}-\varepsilon_{F}\approx
v_{F}(k-k_{F})$ is energy of an electron counted from Fermi
surface, $\varepsilon$ is an energy parameter, $\delta\rightarrow
0$, we use a system of units where $\hbar=k_{\texttt{B}}=1$. State
of an impurity can be described with Green function:
\begin{eqnarray}\label{1.3}
\mathcal{G}_{A}(\varepsilon)=\frac{1}{\gamma-E_{A}+i\delta}
\end{eqnarray}
A system described with Hamiltonian (\ref{1.1}) is nonhomogeneous
and momentums of quasiparticles are not conserved. However
averaging over an ensemble of samples with all possible positions
of impurities recovers spatial homogeneity of the system, and
quasiparticles' momentums are conserved (Appendix \ref{A}). The
averaging operation over a disorder has a form \citep{levit}:
\begin{eqnarray}\label{1.4}
\left\langle
G(x,x')\right\rangle=-i\left\langle\frac{\left\langle\widehat{T}\psi^{+}(x)\psi(x')\widehat{U}\right\rangle_{0}}
{\left\langle\widehat{U}\right\rangle_{0}}\right\rangle_{\texttt{disorder}},
\end{eqnarray}
where $\widehat{U}$ is an evolution operator,
$\langle\ldots\rangle_{0}$ is done over a ground state of Fermi
system and a lattice (in the numerator and the denominator
separately). The averaging over the disorder is done in the
following way - at first the propagator is calculated at the given
disorder, and only then the averaging $\langle\ldots\rangle$ is
done (the whole fraction is averaged). Practically the averaging
$\left\langle\right\rangle_{\texttt{disorder}}$ is done as
follows:
\begin{equation}\label{1.4a}
\sum_{j}\int d\textbf{r}_{j} \longrightarrow\frac{N}{V}\int \int
d\textbf{r}' d\textbf{R}
\end{equation}
Conservation of momentum allows us to summarize diagrams with help
of Dyson equation (\ref{A.5}).

\begin{figure}[ht]
\includegraphics[width=4.0cm]{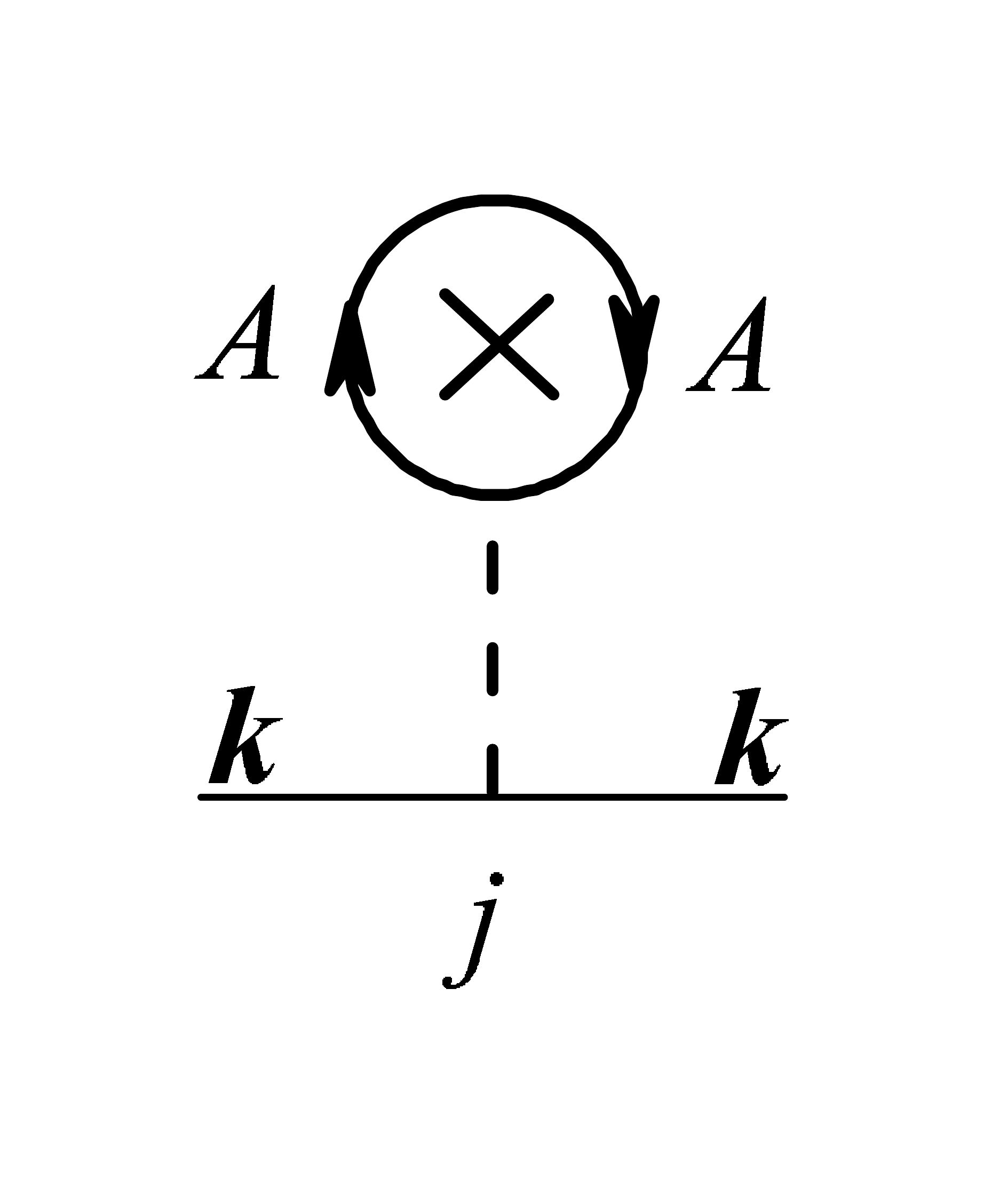}
\caption{The fist order diagram describing a retarded scattering
of a quasiparticle by impurities. The impurity does not change its
state during the process $|A\rangle\rightarrow|A\rangle$.}
\label{Fig2}
\end{figure}
Let us consider the simplest process which is analogous to a
process in Fig.{\ref{Fig1A}}. The process is represented in
Fig.{\ref{Fig2}}. It means an electron interacts with an impurity
no changing impurity's state ($\phi_{A}\rightarrow\phi_{A}$). That
is the scattering is elastic. In analytical form a mass operator
is
\begin{eqnarray}\label{1.5}
&&(-i)\Sigma=\int_{-\infty}^{+\infty}
i\mathcal{G}_{A}(\gamma)\frac{d\gamma}{2\pi}(-i)\sum_{j}\int\int\phi_{A}^{+}
\left(\textbf{r}_{j}-\textbf{R}_{j}\right)\psi_{\textbf{k}}^{+}(\textbf{r})
U(\textbf{r}-\textbf{r}_{j})\phi_{A}\left(\textbf{r}_{j}-\textbf{R}_{j}\right)\psi_{\textbf{k}}(\textbf{r})d\textbf{r}d\textbf{r}_{j}\nonumber\\
&&=-i\frac{N}{V^{2}}\int\int\int
U(\textbf{r}-\textbf{r}')\left|\phi_{A}(\textbf{r}'-\textbf{R})\right|^{2}d\textbf{r}d\textbf{r}'d\textbf{R}\nonumber\\
&&=-i\frac{N}{V}\int\int
U(\textbf{r}-\textbf{r}')\left|\phi_{A}(\textbf{r}'-\textbf{R})\right|^{2}d(\textbf{r}-\textbf{r}')d(\textbf{r}'-\textbf{R})=-i\frac{N}{V}\int
U(R)d\textbf{R}=-i\rho U(q=0),
\end{eqnarray}
where $\phi_{A}\equiv|A\rangle$ is a ground state of the impurity
$\left\langle A|A\right\rangle=1$, and we took advantage in that
integration over $\textbf{r}$, $\textbf{r}'$ and $\textbf{R}$ is
done over infinite volume $\int
d\textbf{r}\equiv\int^{+\infty}_{-\infty}r^{2}dr\int^{\pi}_{0}\sin\theta
d\theta\int^{2\pi}_{0}d\varphi$, $\int d\textbf{R}=V$,
$N\rightarrow\infty,V\rightarrow\infty,N/V=\rho=\texttt{const}$.
Contribution of this diagram is trivial - it shifts a chemical
potential only: $\mu-\rho U(0)$.

In a case of nonzero temperature $T\neq 0$ the impurities are
distributed over states $|A\rangle,|B\rangle,|C\rangle,\ldots$
with probability
\begin{equation}\label{1.6}
    \varpi_{A}=\frac{1}{Z}\exp\left(-\frac{E_{A}-E_{0}}{T}\right),\quad Z=\sum_{A}\exp\left(-\frac{E_{A}-E_{0}}{T}\right)
\end{equation}
where $E_{0}$ is an energy of ground state of an impurity, and the
summation is extended on all possible states (we use a system of
units where $\hbar=k_{\texttt{B}}=1$). Then
\begin{eqnarray}\label{1.7}
&&-\Sigma=-i\rho\sum_{A}\varpi_{A}\int\int
U(\textbf{r}-\textbf{r}')\left|\phi_{A}(\textbf{r}'-\textbf{R})\right|^{2}d(\textbf{r}-\textbf{r}')d(\textbf{r}'-\textbf{R})=-\rho
U(q=0),
\end{eqnarray}
because $\sum_{A}\varpi_{A}=1$, $\left\langle A|A\right\rangle=1$.

\begin{figure}[h]
\includegraphics[width=12cm]{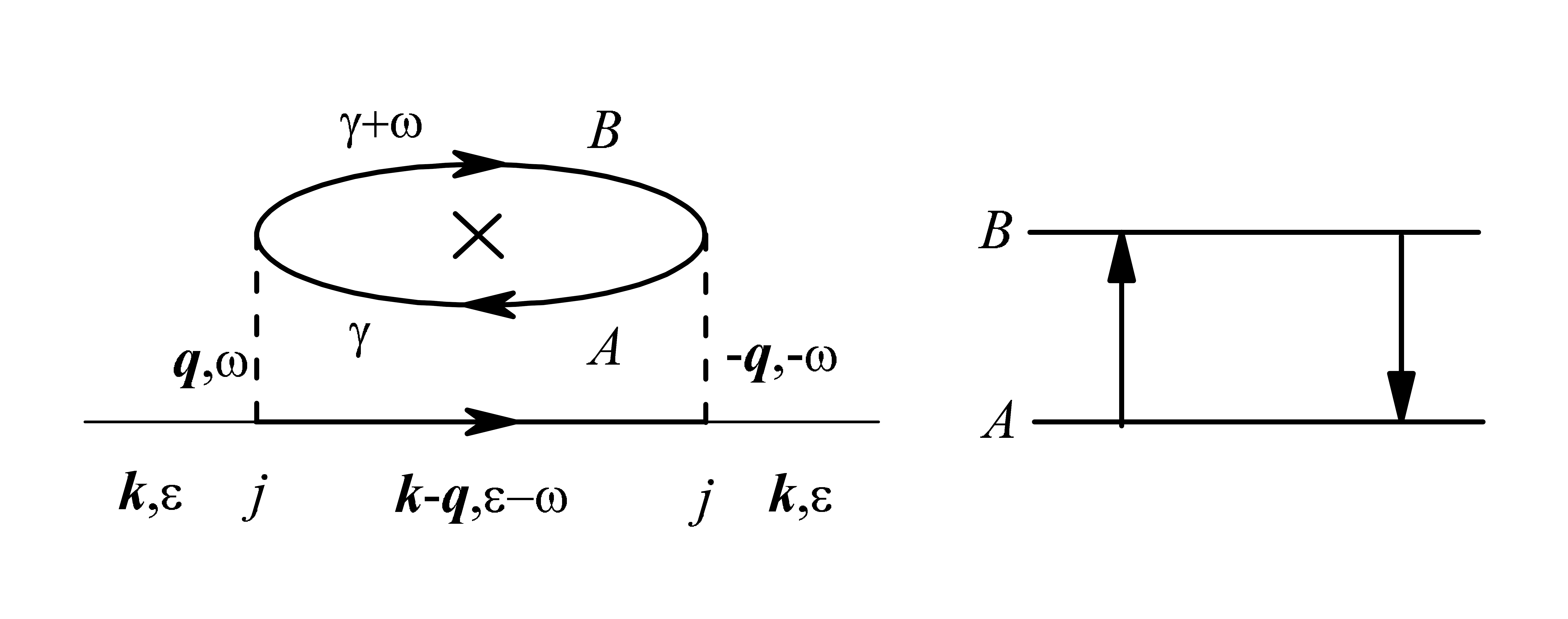}
\caption{The second order diagram describing a retarded
interaction of a quasiparticle by impurities. The impurity's state
is changed during the process
$|A\rangle\rightarrow|B\rangle\rightarrow|A\rangle$ (a right-hand
picture).} \label{Fig3}
\end{figure}

Let us consider the second order process represented in
Fig.\ref{Fig3}. It is analogous to the second order process of
elastic scattering shown in Fig.\ref{Fig2A}a. In the inelastic
process an electron interacts with an impurity changing impurity's
state $\phi_{A}\rightarrow\phi_{B}\rightarrow\phi_{A}$. That is a
state $|B\rangle$ is virtual, a state $|A\rangle$ is a ground
state (for $T=0$ only). A transition frequency is
$\omega_{AB}=E_{B}-E_{A}$. This process means a retarded
interaction with impurities. The interaction was considered in the
simplest form in an article \cite{grig}.
In an analytical form a mass operator is
\begin{eqnarray}\label{1.8}
&&(-i)\Sigma=\sum_{j}\sum_{B}V\int
\frac{d\textbf{q}}{(2\pi)^{3}}\int_{-\infty}^{+\infty}
\frac{d\omega}{2\pi}\int_{-\infty}^{+\infty}
\left[i\mathcal{G}_{A}^{-}(\gamma)i\mathcal{G}_{B}^{+}(\gamma+\omega)+i\mathcal{G}_{A}^{-}(\gamma)i\mathcal{G}_{B}^{+}(\gamma-\omega)\right]\frac{d\gamma}{2\pi}\nonumber\\
&&(-i)\int\int\phi_{B}^{+}\left(\textbf{r}_{j}-\textbf{R}_{j}\right)\psi_{\textbf{k}-\textbf{q}}^{+}(\textbf{r})
U(\textbf{r}-\textbf{r}_{j})\phi_{A}\left(\textbf{r}_{j}-\textbf{R}_{j}\right)\psi_{\textbf{k}}(\textbf{r})d\textbf{r}d\textbf{r}_{j}\nonumber\\
&&(-i)\int\int\phi_{A}^{+}\left(\textbf{r}_{j}-\textbf{R}_{j}\right)\psi_{\textbf{k}}^{+}(\textbf{r})
U(\textbf{r}-\textbf{r}_{j})\phi_{B}\left(\textbf{r}_{j}-\textbf{R}_{j}\right)\psi_{\textbf{k}-\textbf{q}}(\textbf{r})d\textbf{r}d\textbf{r}_{j}\nonumber\\
&&iG_{0}(\textbf{k}-\textbf{q},\varepsilon-\omega)
\end{eqnarray}
The term in square brackets describes virtual transitions of an
impurity between levels $A$ and $B$. If an impurity's level is
empty then its propagator is:
\begin{eqnarray}\label{1.9}
\mathcal{G}_{A}^{-}(\gamma)=-\frac{1}{\gamma-E_{A}-i\delta},
\end{eqnarray}
because the empty level equivalent to a propagation back in time
(a hole or an antiparticle), and many impurities in a system can
be in the same state, that is the minus sign must be both for
bosons. An electron interacting with an impurity takes it to a
state $\phi_{B}$ and gives it an energy parameter $\omega$. As
Feynman diagrams are not ordered in time \cite{matt} then
Eq.(\ref{1.8}) is a sum of two terms: with the positive and
negative energy parameter:
\begin{eqnarray}\label{1.10}
&&\int_{-\infty}^{+\infty}
\left[i\mathcal{G}_{A}^{-}(\gamma)i\mathcal{G}_{B}^{+}(\gamma+\omega)+i\mathcal{G}_{A}^{-}(\gamma)i\mathcal{G}_{B}^{+}(\gamma-\omega)\right]
\frac{d\gamma}{2\pi}\nonumber\\
&&=i\frac{2\omega_{AB}}{\omega^{2}-\omega_{AB}^{2}+i2\omega_{AB}\delta}\equiv
iD_{AB}(\omega).
\end{eqnarray}
A function $D_{AB}(\omega)$ has a form of a collective
excitation's propagator. Therefore we name this function as
"pseudopropagator" \cite{grig}. A prefix "pseudo" means this
propagator is a result of the averaging (\ref{1.4}) and the
corresponding correlations (\ref{A.4}).

Now let us consider an integral:
\begin{eqnarray}
\sum_{j}&&\int\int\phi_{B}^{+}\left(\textbf{r}_{j}-\textbf{R}_{j}\right)\psi_{\textbf{k}-\textbf{q}}^{+}(\textbf{r})
U(\textbf{r}-\textbf{r}_{j})\phi_{A}\left(\textbf{r}_{j}-\textbf{R}_{j}\right)\psi_{\textbf{k}}(\textbf{r})d\textbf{r}d\textbf{r}_{j}\nonumber\\
&&\int\int\phi_{A}^{+}\left(\textbf{r}_{j}-\textbf{R}_{j}\right)\psi_{\textbf{k}}^{+}(\textbf{r})
U(\textbf{r}-\textbf{r}_{j})\phi_{B}\left(\textbf{r}_{j}-\textbf{R}_{j}\right)\psi_{\textbf{k}-\textbf{q}}(\textbf{r})d\textbf{r}d\textbf{r}_{j}\nonumber\\
=\frac{N}{V^{3}}\int d\textbf{R}&&\int\int e^{i\textbf{qr}}
U(\textbf{r}-\textbf{r}')\phi_{B}^{+}\left(\textbf{r}'-\textbf{R}\right)\phi_{A}\left(\textbf{r}'-\textbf{R}\right)d\textbf{r}d\textbf{r}'\nonumber\\
&&\int\int e^{-i\textbf{qr}}
U(\textbf{r}-\textbf{r}')\phi_{A}^{+}\left(\textbf{r}'-\textbf{R}\right)\phi_{B}\left(\textbf{r}'-\textbf{R}\right)d\textbf{r}d\textbf{r}'\nonumber\\
=\frac{N}{V^{3}}\int d\textbf{R}&&\int\int
e^{i\textbf{q}\left(\textbf{r}-\textbf{r}'\right)}
U(\textbf{r}-\textbf{r}')e^{i\textbf{q}\left(\textbf{r}'-\textbf{R}\right)}\phi_{B}^{+}\left(\textbf{r}'-\textbf{R}\right)\phi_{A}\left(\textbf{r}'-\textbf{R}\right)
d\left(\textbf{r}-\textbf{r}'\right)d\left(\textbf{r}'-\textbf{R}\right)\nonumber\\
&&\int\int e^{-i\textbf{q}\left(\textbf{r}-\textbf{r}'\right)}
U(\textbf{r}-\textbf{r}')e^{-i\textbf{q}\left(\textbf{r}'-\textbf{R}\right)}\phi_{A}^{+}\left(\textbf{r}'-\textbf{R}\right)\phi_{B}\left(\textbf{r}'-\textbf{R}\right)
d\left(\textbf{r}-\textbf{r}'\right)d\left(\textbf{r}'-\textbf{R}\right)\nonumber\\
&&=\frac{\rho}{V}U(\textbf{q})U(-\textbf{q})\langle
B|A\rangle_{\textbf{q}}\langle
A|B\rangle_{-\textbf{q}}=\frac{\rho}{V}\left|U(\textbf{q})\langle
B|A\rangle_{\textbf{q}}\right|^{2},\nonumber
\end{eqnarray}
where
\begin{eqnarray}\label{1.11}
U(\textbf{q})&=&\int
e^{i\textbf{q}\textbf{r}}U(\textbf{r})d\textbf{r} \\
\langle B|A\rangle_{\textbf{q}}&=&\int
e^{i\textbf{q}\textbf{r}}\phi_{B}^{+}(\textbf{r})\phi_{A}(\textbf{r})d\textbf{r}.\label{1.12}
\end{eqnarray}
Then a mass operator is
\begin{eqnarray}\label{1.13}
-i\Sigma(\textbf{k},\varepsilon)=\rho\sum_{B}\int\frac{d\textbf{q}d\omega}{(2\pi)^{4}}\left|U(\textbf{q})\langle
B|A\rangle_{\textbf{q}}\right|^{2}(-i)D_{AB}(\omega)iG_{0}(\textbf{k}-\textbf{q},\varepsilon-\omega)
\end{eqnarray}
We can see this expression for a mass operator is analogous to
electron-phonon interaction (phonons with Einstein specter), where
a value $\rho\left|U(\textbf{q})\langle
B|A\rangle_{\textbf{q}}\right|^{2}$ plays a role of a coupling
constant. Eq.\ref{1.13} corresponds to result of an article
\cite{grig} (if we suppose $\left|U(\textbf{q})\langle
B|A\rangle_{\textbf{q}}\right|^{2}=\texttt{const}$ and an impurity
is two-level system) when the integration over $q$ be done within
the boundaries $\int_{0}^{2k_{F}}q^{2}dq$. We have to consider a
case when $\omega_{AB}=0, \phi_{B}=\phi_{A}$, that is the
scattering is elastic. Then energetic parameter $\omega$ is not
transferred along the line of interaction. Then according to the
rules of a diagram technique an integration over the intermediate
energetic parameter is absent. Instead of Eq.(\ref{1.10}) we must
have $\int_{-\infty}^{+\infty}
i\mathcal{G}_{A}^{+}(\gamma)\frac{d\gamma}{2\pi}=1$. Then we have
\begin{eqnarray}\label{1.13a}
-i\Sigma(\textbf{k},\varepsilon,\omega_{AB}=0)=\rho\int\frac{d\textbf{q}d\omega}{(2\pi)^{4}}(-i)U(\textbf{q})(-i)U(-\textbf{q})\left|\langle
A|A\rangle_{\textbf{q}}\right|^{2}iG_{0}(\textbf{k}-\textbf{q},\varepsilon).
\end{eqnarray}
Impurity's ground state wave function have a form $\phi_{A}\sim
e^{-r/a}$. If $1/a\gg k_{F}$ then we can assume $\langle
A|A\rangle_{\textbf{q}}\approx\langle A|A\rangle_{0}=1$. Then
Eq.(\ref{1.13a}) coincides with Eq.(\ref{A.6}) for a mass operator
of the second order elastic process.

Let us generalize Eq.(\ref{1.13}) for nonzero temperatures and do
some transformations. Following \cite{mahan} let us denote
$p^{2}\equiv|\textbf{k}-\textbf{q}|^{2}=k^{2}+q^{2}+2kq\cos\theta$,
where $\theta=\widehat{\textbf{k},-\textbf{q}}$ and $k\approx
k_{F}$. Further we have $pdp=kqd(\cos\theta)$,
$\xi=v_{F}(p-p_{F})\Rightarrow d\xi=v_{F}dp$. Then $pdp=md\xi$.
Hence
\begin{eqnarray}
\int d\textbf{q}\equiv \int
d^{3}q=\int_{0}^{2k_{F}}q^{2}dq\int^{1}_{-1}
d(\cos\theta)\int^{2\pi}_{0}d\varphi=\frac{2\pi}{k_{F}}\int_{0}^{2k_{F}}
qdq\int pdp=\int\frac{d^{2}q}{v_{F}}\int_{-\infty}^{\infty}
d\xi,\nonumber
\end{eqnarray}
where limits of integration over $d\xi$ can be extended between
$\pm\infty$ because the main contribution of the integrand is in
the region $\xi\approx 0$. Then we can write
\begin{eqnarray}\label{1.14}
-\Sigma(\textbf{k},\varepsilon_{n})=\rho
T\sum_{A}\sum_{B}\varpi_{A}\sum_{m=-\infty}^{+\infty}\int\int\frac{d^{2}qd\xi}{v_{F}(2\pi)^{3}}\left|U(\textbf{q})\langle
B|A\rangle_{\textbf{q}}\right|^{2}iD_{AB}(\varepsilon_{n}-\varepsilon_{m})iG_{0}(\xi,\varepsilon_{m}),
\end{eqnarray}
where an electron's propagator and a pseudopropagator are
\begin{eqnarray}\label{1.15}
G_{0}(\xi,\varepsilon_{m})=-i\frac{i\varepsilon_{m}+\xi}{\varepsilon_{m}^{2}+\xi^{2}},\quad
D_{AB}(\varepsilon_{n}-\varepsilon_{m})=-i\frac{2\omega_{AB}}{(\varepsilon_{n}-\varepsilon_{m})^{2}+\omega_{AB}^{2}}.
\end{eqnarray}
and $\varepsilon_{m}=\pi T(2m+1)$. It should be noted an
interaction function $U(\textbf{q})$ can be represented via a
differential scattering cross-section:
$\frac{d\sigma}{d\Omega}=\left|\frac{m}{2\pi}U(\textbf{q})\right|^{2}$.

\begin{figure}[h]
\includegraphics[width=12cm]{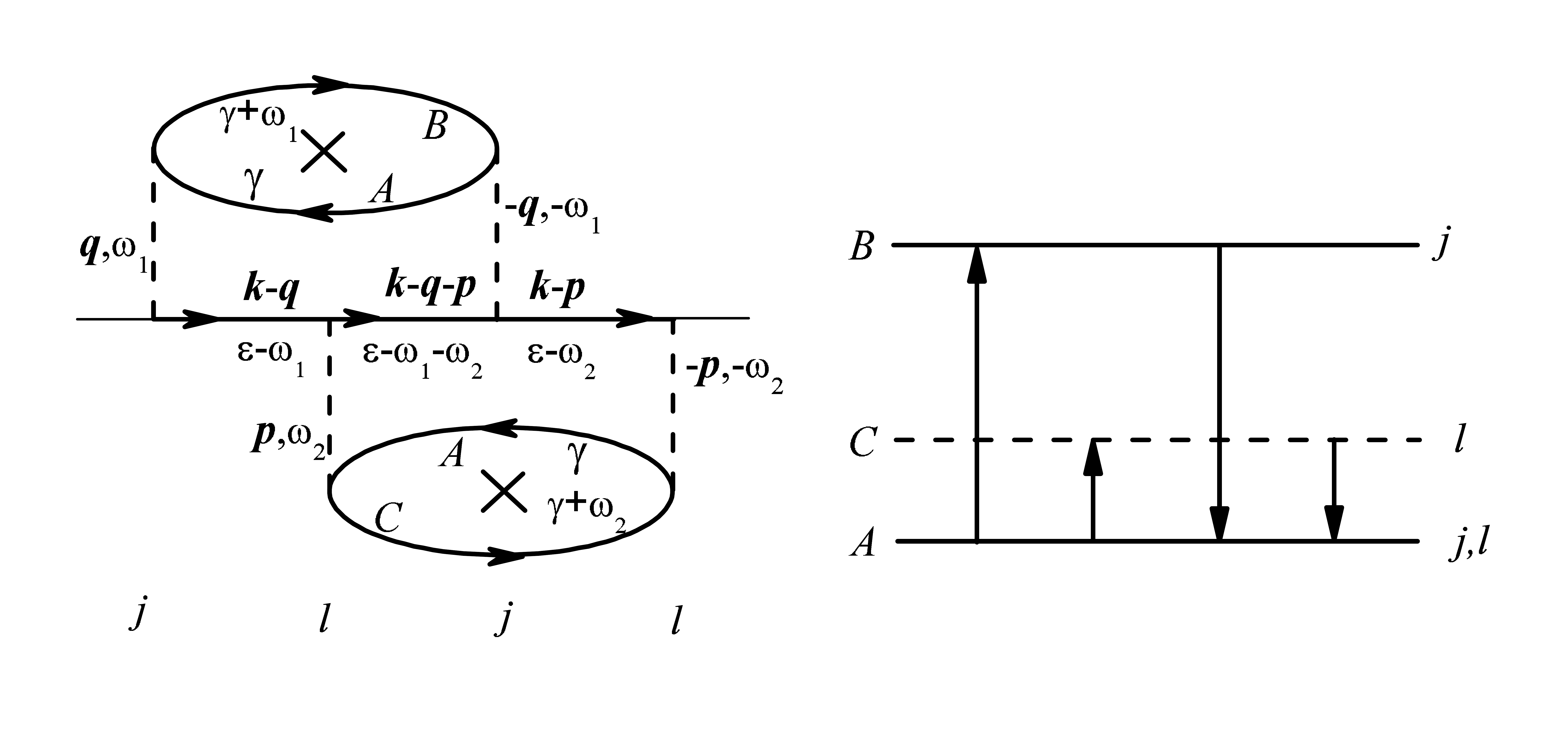}
\caption{The fourth order cross-diagram. In the scattering two
impurities ($j$th and $l$th) take part simultaneously. The
impurities' states are changed during the process: $j$:
$|A\rangle\rightarrow|B\rangle\rightarrow|A\rangle$, $l$:
$|A\rangle\rightarrow|C\rangle\rightarrow|A\rangle$  (a right-hand
picture).} \label{Fig4}
\end{figure}

Higher order diagrams may be classified under two types. The first
type corresponds to cross-diagrams. They describe processes like
discussed above second order process, however the scattering takes
place by different impurities. The simplest fourth order diagram
is shown in Fig.\ref{Fig4}. The diagram means an electron
interacts with an impurity $j$ changing its state
$\phi_{A}\rightarrow\phi_{B}$, then the electron interacts with an
impurity $l$ changing its state $\phi_{A}\rightarrow\phi_{C}$.
Then the electron gathers the energies interacting again with the
impurities $\phi_{B}\rightarrow\phi_{A}$,
$\phi_{C}\rightarrow\phi_{A}$. States $|B\rangle$ and $|C\rangle$
are virtual, a state $|A\rangle$ is a ground state (for $T=0$
only). Transition frequencies are $\omega_{AB}=|E_{B}-E_{A}|$,
$\omega_{AC}=|E_{C}-E_{A}|$. Analytically the process is
represented as follows:
\begin{eqnarray}\label{1.16}
-i\Sigma(\textbf{k},\varepsilon)=&&\rho^{2}\sum_{B}\sum_{C}\int\frac{d\textbf{q}d\omega_{1}}{(2\pi)^{4}}\int\frac{d\textbf{p}d\omega_{2}}{(2\pi)^{4}}
\left|U(\textbf{q})\langle
B|A\rangle_{\textbf{q}}\right|^{2}\left|U(\textbf{p})\langle
C|A\rangle_{\textbf{p}}\right|^{2}(-i)D_{AB}(\omega_{1})(-i)D_{AC}(\omega_{2})\nonumber\\
&&iG_{0}(\textbf{k}-\textbf{q},\varepsilon-\omega_{1})
iG_{0}(\textbf{k}-\textbf{p},\varepsilon-\omega_{2})iG_{0}(\textbf{k}-\textbf{q}-\textbf{p},\varepsilon-\omega_{1}-\omega_{2}).
\end{eqnarray}
Analogously we can consider any cross-diagrams of higher orders.

\begin{figure}[h]
\includegraphics[width=8cm]{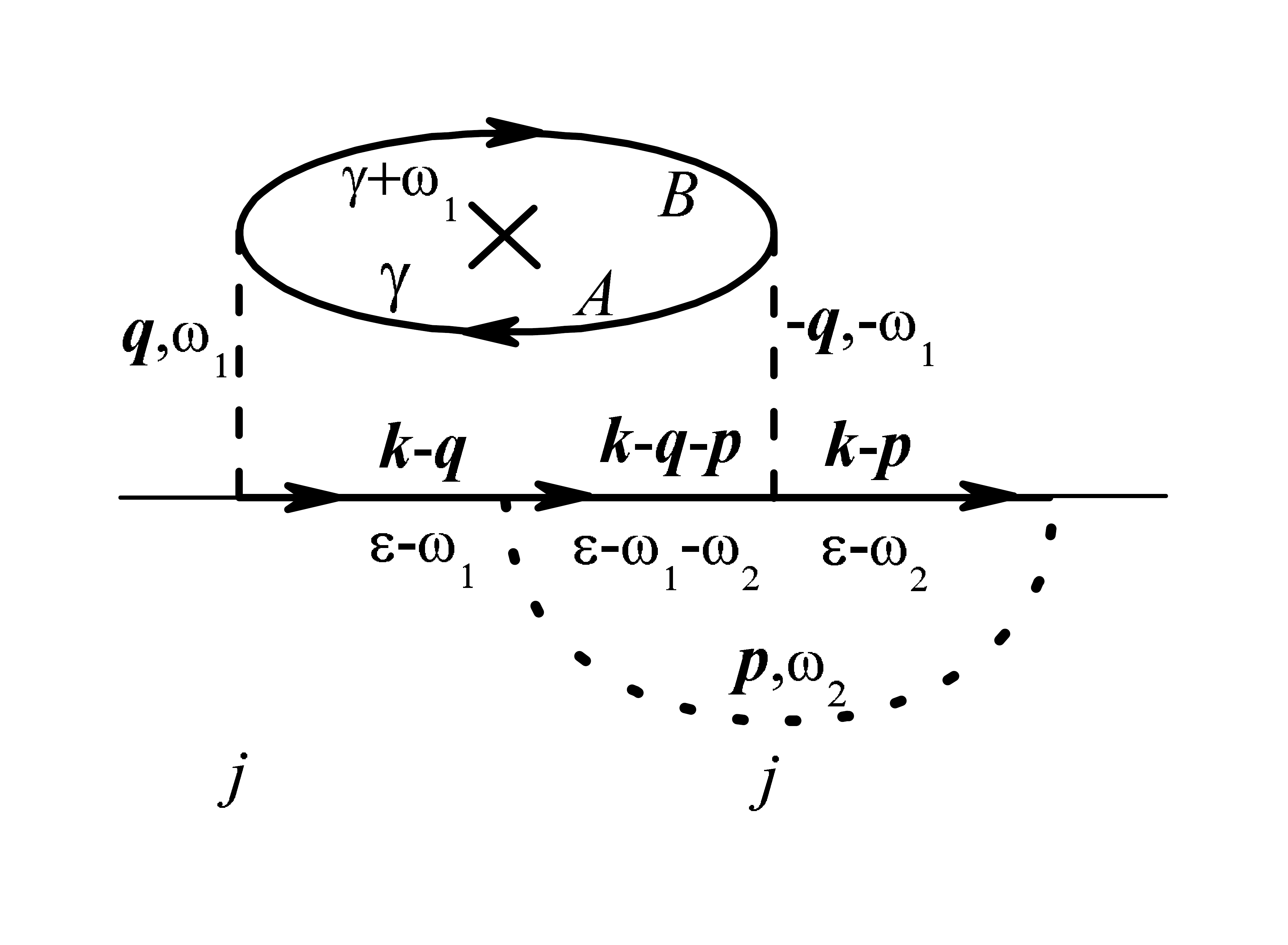}
\caption{The fourth order cross-diagram with involvement of a
phonon. In scattering one impurities and one phonon take place
simultaneously.} \label{Fig5}
\end{figure}

A  next type of cross-diagrams is shown in Fig.\ref{Fig5}. In this
diagram a scattering process by an impurity is crossed with an
electron-phonon interaction. Analytically the process is
represented as follows:
\begin{eqnarray}\label{1.17}
-i\Sigma(\textbf{k},\varepsilon)=&&\rho\sum_{B}\int\frac{d\textbf{q}d\omega_{1}}{(2\pi)^{4}}\int\frac{d\textbf{p}d\omega_{2}}{(2\pi)^{4}}
\left|U(\textbf{q})\langle
B|A\rangle_{\textbf{q}}\right|^{2}\left|g(\textbf{p})\right|^{2}(-i)D_{AB}(\omega_{1})(-i)D_{\texttt{ph}}(\omega_{2},\Omega(\textbf{p}))\nonumber\\
&&iG_{0}(\textbf{k}-\textbf{q},\varepsilon-\omega_{1})
iG_{0}(\textbf{k}-\textbf{p},\varepsilon-\omega_{2})iG_{0}(\textbf{k}-\textbf{q}-\textbf{p},\varepsilon-\omega_{1}-\omega_{2}),
\end{eqnarray}
where $g(\textbf{p})$ is an electron-phonon coupling constant,
$D_{\texttt{ph}}(\omega_{2},\Omega(\textbf{p}))$ is a phonon
propagator.

A small parameter for the expansion is a ratio of a contribution
of cross-diagrams to a contribution of diagrams without crossings.
In \citep{sad} it was shown that the ratio is proportional to
$\frac{\triangle k}{k_{F}}$, where $\triangle k$ is a momentum's
uncertainty as result of scattering $\triangle k\propto 1/l$ ($l$
is a free length). Then the small parameter is $1/lk_{F}\ll 1$,
that is correct for a weak coupling. At inelastic scattering by
impurities a particle's energy change by a value
$\triangle\varepsilon\sim\omega_{AB}$, that corresponds to a
momentum's uncertainty $\triangle k=\frac{m\omega_{AB}}{k_{F}}$.
Hence the small parameter is
\begin{equation}\label{1.18}
\frac{\triangle
k}{k_{F}}=\frac{m\omega_{AB}}{k_{F}^{2}}\sim\frac{\omega_{AB}}{\varepsilon_{F}}\ll
1.
\end{equation}
And so on for each frequency $\omega_{AB},\omega_{CD},\ldots$.
Eq.\ref{1.18} likes a situation with phonons where a small
parameter is an adiabaticity parameter (Migdal's theorem).

\begin{figure}[h]
\includegraphics[width=12cm]{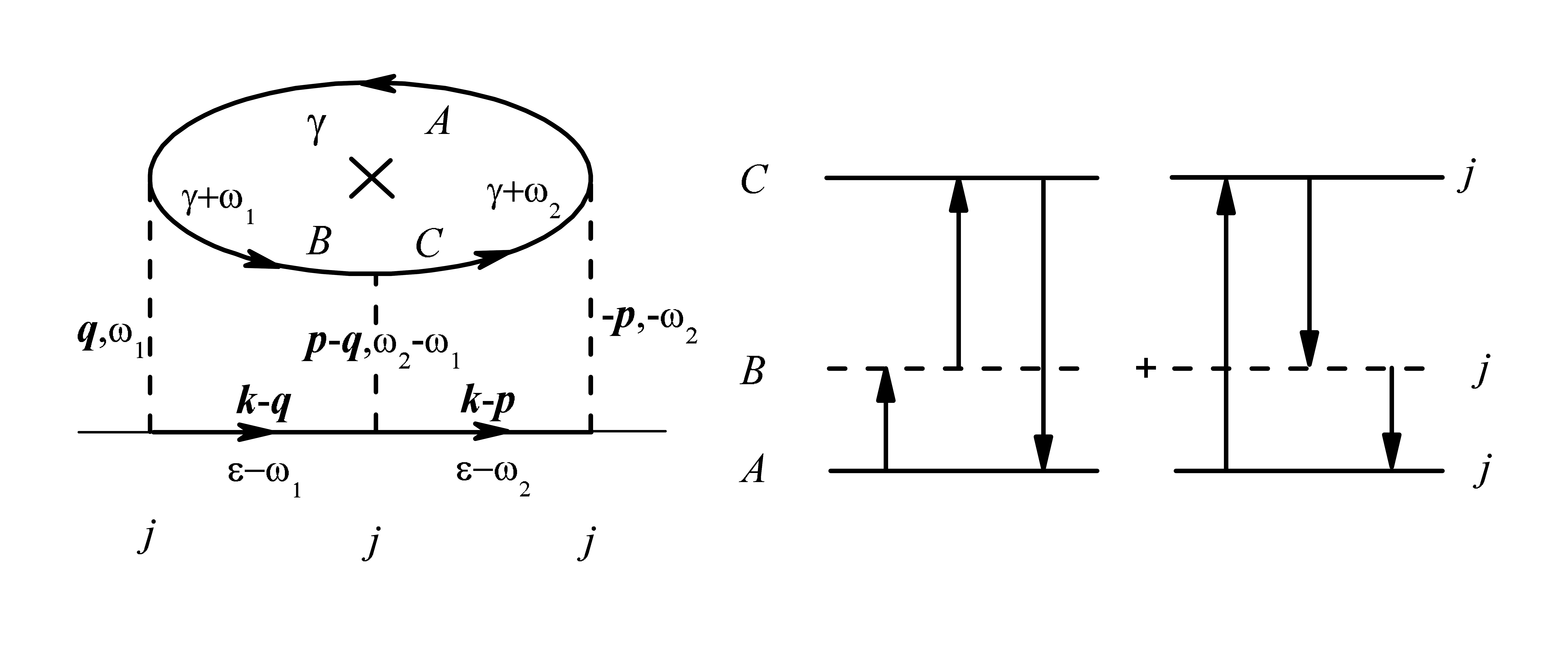}
\caption{The third order beam-like diagram. In scattering one
impurity takes part only, but it must have three level as minimum.
The impurity's state is changed during the process
$|A\rangle\rightarrow|B\rangle\rightarrow|C\rangle\rightarrow|A\rangle$
in two ways (a right-hand picture).} \label{Fig6}
\end{figure}

The second type corresponds to beam-like diagrams. For elastic
scattering the diagrams is shown in Fig.\ref{Fig2A}c. For
inelastic scattering a beam-type diagram of third order is shown
in Fig.\ref{Fig6}.
\begin{figure}[ht]
\includegraphics[width=8cm]{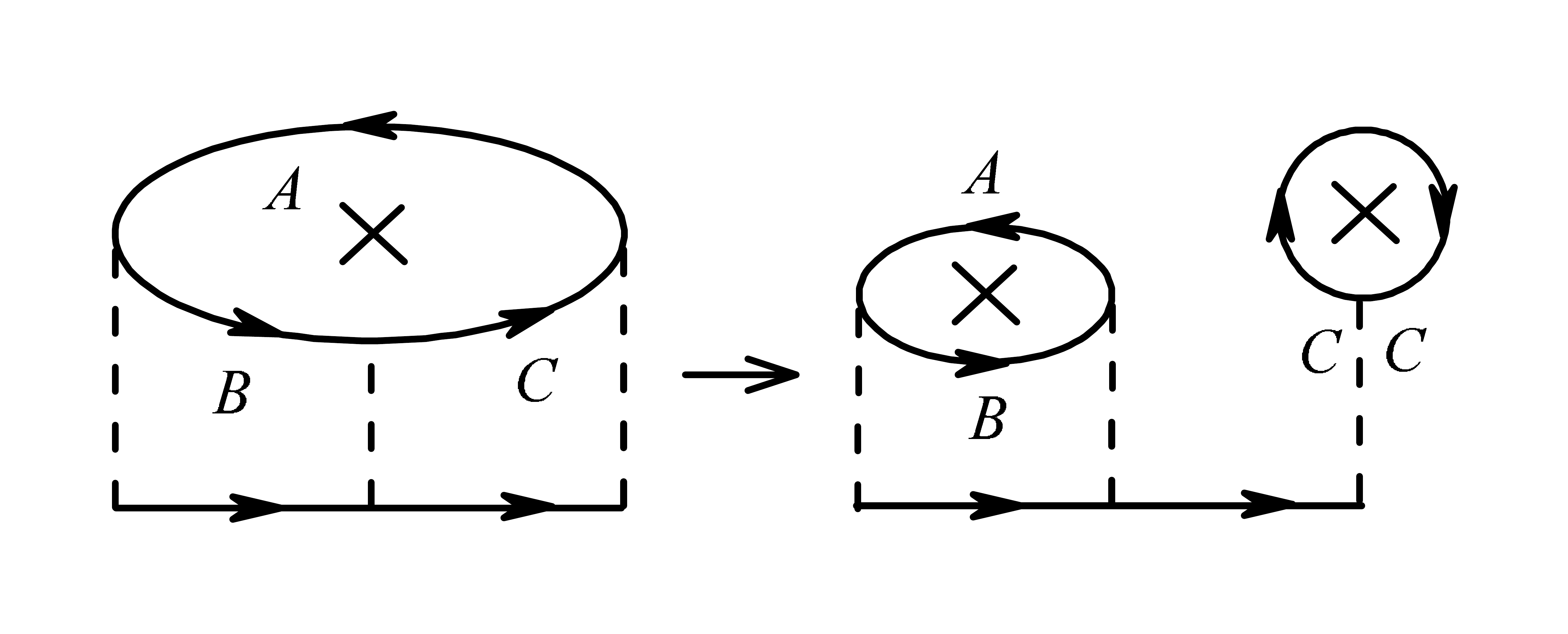}
\caption{Uncoupling of a beam-like third-order diagram to a
reducible diagram containing the second order part and the first
order part.} \label{Fig7}
\end{figure}
The diagram means an electron interacts with an impurity $j$
changing its state $\phi_{A}\rightarrow\phi_{B}$. Then the
electron interacts with the impurity again changing a impurity's
state $\phi_{B}\rightarrow\phi_{C}$. Then the electron recovers
energy interacting again with the impurity
$\phi_{C}\rightarrow\phi_{A}$. States $|B\rangle$ and $|C\rangle$
are virtual, a state $|A\rangle$ is a ground state (for $T=0$
only). Transition frequencies are $\omega_{AB}=E_{B}-E_{A}$,
$\omega_{AC}=E_{C}-E_{A}$, $\omega_{BC}=E_{C}-E_{B}$.  In addition
another variant of the process is possible
$|A\rangle\rightarrow|C\rangle\rightarrow|B\rangle\rightarrow|A\rangle$.
Analytically the process is represented as follows:
\begin{eqnarray}\label{1.19}
-i\Sigma(\textbf{k},\varepsilon)=&&\rho\sum_{B}\sum_{C}\int\frac{d\textbf{q}d\omega_{1}}{(2\pi)^{4}}\int\frac{d\textbf{p}d\omega_{2}}{(2\pi)^{4}}
U(\textbf{q})\langle B|A\rangle_{\textbf{q}}U(\textbf{p}-\textbf{q})\langle C|B\rangle_{\textbf{p}-\textbf{q}}U(-\textbf{p})\langle A|C\rangle_{-\textbf{p}}\nonumber\\
&&iG_{0}(\textbf{k}-\textbf{q},\varepsilon-\omega_{1})
iG_{0}(\textbf{k}-\textbf{p},\varepsilon-\omega_{2})\nonumber\\
&&\int\frac{d\gamma}{2\pi}[i\mathcal{G}_{A}^{-}(\gamma)i\mathcal{G}_{B}^{+}(\gamma+\omega_{1})i\mathcal{G}_{C}^{+}(\gamma+\omega_{2})
+i\mathcal{G}_{A}^{-}(\gamma)i\mathcal{G}_{B}^{+}(\gamma-\omega_{1})i\mathcal{G}_{C}^{+}(\gamma-\omega_{2})\nonumber\\
&&i\mathcal{G}_{A}^{-}(\gamma)i\mathcal{G}_{B}^{+}(\gamma-\omega_{1})i\mathcal{G}_{C}^{+}(\gamma+\omega_{2})+i\mathcal{G}_{A}^{-}(\gamma)i\mathcal{G}_{B}^{+}(\gamma+\omega_{1})i\mathcal{G}_{C}^{+}(\gamma-\omega_{2})]\nonumber\\
=&&\rho\sum_{B}\sum_{C}\int\frac{d\textbf{q}d\omega_{1}}{(2\pi)^{4}}\int\frac{d\textbf{p}d\omega_{2}}{(2\pi)^{4}}
U(\textbf{q})\langle B|A\rangle_{\textbf{q}}U(\textbf{p}-\textbf{q})\langle C|B\rangle_{\textbf{p}-\textbf{q}}U(-\textbf{p})\langle A|C\rangle_{-\textbf{p}}\nonumber\\
&&iG_{0}(\textbf{k}-\textbf{q},\varepsilon-\omega_{1})
iG_{0}(\textbf{k}-\textbf{p},\varepsilon-\omega_{2})\left[\frac{-4\omega_{AB}\omega_{AC}}{(\omega_{1}^{2}-\omega_{AB}^{2})(\omega_{2}^{2}-\omega_{AC}^{2})}\right].
\end{eqnarray}
Contribution of this process is proportional to $\rho U^{3}$.
Analogously we can construct higher order diagrams of this type to
be proportional to $\rho U^{4},\rho U^{5},\ldots$. The beam-like
diagrams violate the analogy with electron-phonon interaction. In
a limit $\rho\rightarrow\infty,U^{2}\rightarrow 0,\rho
U^{2}=\texttt{const}$ (continuous "spreading" of impurities over a
system) the diagrams disappear. Unfortunately the series of
beam-like diagrams have not any small parameter like the series of
cross-diagrams. However it is not difficult to notice that the
mass operator of the third order beam-like process is
$\Sigma_{ABC}\propto\rho\int
d\textbf{p}U(\textbf{q})U(\textbf{p}-\textbf{q})U(-\textbf{p})\langle
B|A\rangle_{\textbf{q}}\langle
C|B\rangle_{\textbf{p}-\textbf{q}}\langle
A|C\rangle_{-\textbf{p}}$ and a mass operator for the third order
process when scattering takes place by different impurities (it is
an reducible diagram
$-i\Sigma_{AB[CC]}=-i\Sigma_{AB}(-i)G_{0}(-i)\Sigma_{CC}$) is
$\Sigma_{AB[CC]}\propto\rho^{2} U(0)|U(\textbf{q})\langle
B|A\rangle_{\textbf{q}}|^{2}\langle C|C\rangle_{0}$. In the
expression $\Sigma_{ABC}$ the integrand is an alternating
function, and in the expression $\Sigma_{AB[CC]}$ the integrand is
a constant-sign function since $\langle B|A\rangle_{0}=0,\langle
C|A\rangle_{0}=0, \langle C|C\rangle_{0}=1$. Therefore due the
integration we have $\Sigma_{ABC}\ll\Sigma_{AB[CC]}$. For higher
orders the alternating is strengthened. Thus we can uncouple
beam-like diagrams as shown in Fig.\ref{Fig7} and thereby
reconstitute the phonon analogy.

\section{Superconducting state.} \label{super}

In this section we generalize results obtained in a previous
section in two directions: to make the perturbation theory as
self-consistent and to apply it for a superconductive state. To
make the perturbation theory as self-consistent the free
propagators $G_{0}$ must be replaced with dressed propagators $G$
(internal lines in diagrams are bolded). To use the perturbation
theory for superconductive state we have to consider anomalous
propagators $F$ and $F^{+}$ which are proportional to order
parameters $\Delta$ and $\Delta^{+}$. We suppose the order
parameter is self-averaging:
$\langle\Delta^{2}(\textbf{r})\rangle-\langle\Delta(\textbf{r})\rangle^{2}=0$.
This means to neglect a scattering of Cooper pairs by fluctuations
of the gap. And we suppose singlet $s$-wave pairing takes place.

\begin{figure}[ht]
\includegraphics[width=15cm]{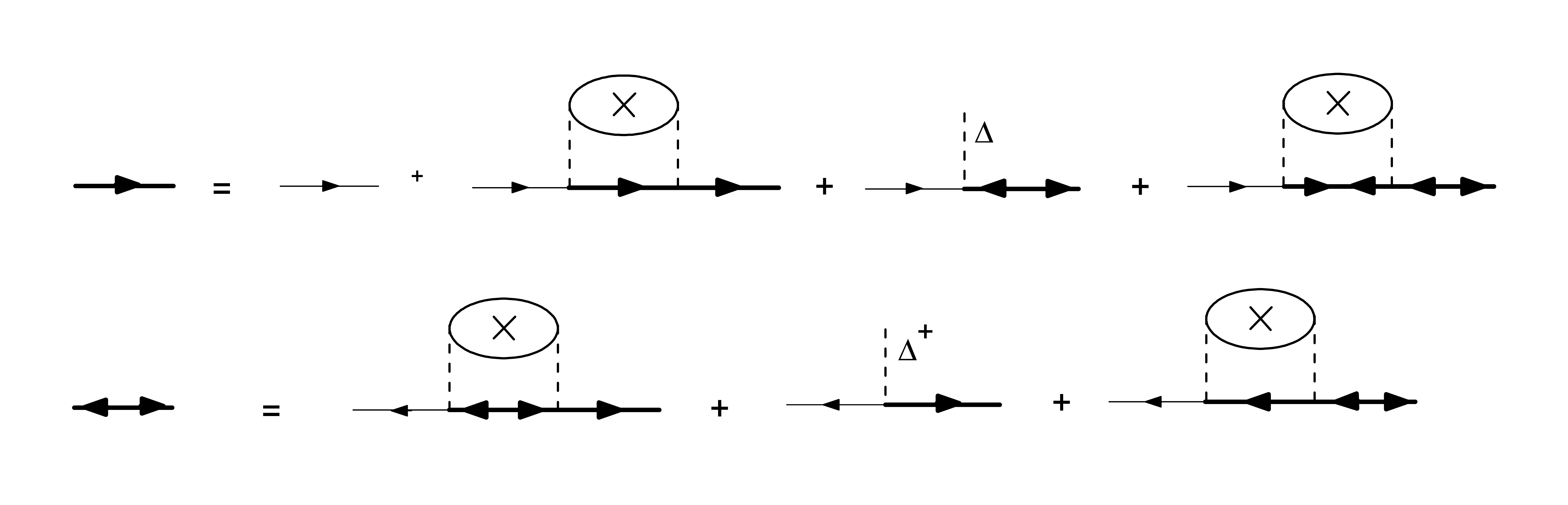}
\caption{Gor'kov equations for a dirty superconductor. Interaction
of quasiparticles with impurities is retarded. Unidirectional thin
lines correspond to free propagators $G_{0}$ and $G_{0}^{+}$ (with
reverse arrows). Unidirectional bold lines correspond to dressed
normal propagators $\widetilde{G}$ and $\widetilde{G}^{+}$. Bold
lines with oppositely directed arrows correspond to dressed
anomalous propagators $\widetilde{F}$ and $\widetilde{F}^{+}$.}
\label{Fig8}
\end{figure}

Gor'kov equations for a dirty superconductor with retarded
interaction of quasiparticles with impurities have a form shown in
Fig.\ref{Fig8}. Their sense is that electrons pair in the metallic
matrix at first, then normal and anomalous propagators are dressed
by interaction with impurities. Unlike elastic interaction the
lines of interaction (dotted lines) transfer energy. The equation
are self-consistent because dressed propagators are calculated
with the dressed propagators (bold lines under interaction with
impurities).
Solutions of the equations are dressed normal and anomalous
propagators:
\begin{equation}\label{2.1}
    \widetilde{G}(\varepsilon_{n},\xi)=-i\frac{i\widetilde{\varepsilon}_{n}+\xi}
    {\widetilde{\varepsilon}_{n}^{2}+\xi^{2}+|\widetilde{\Delta}_{n}|^{2}},\quad
    \widetilde{F}^{+}(\widetilde{\varepsilon}_{n},\xi)=\frac{i\widetilde{\Delta}^{+}_{n}}
    {\widetilde{\varepsilon}_{n}^{2}+\xi^{2}+|\widetilde{\Delta}_{n}|^{2}},
\end{equation}
where a renormalized gap $\widetilde{\Delta}$ and a renormalized
energy parameter $\widetilde{\varepsilon}_{n}$ are determined with
equations (here $\varepsilon_{n}=(2n+1)\pi T$ and
$\varepsilon_{m}=(2m+1)\pi T$):
\begin{eqnarray}
&&\widetilde{\Delta}_{n}=\Delta_{n}+\rho
T\sum_{A}\sum_{B}\varpi_{A}\sum_{m=-\infty}^{+\infty}\int\int\frac{d^{2}qd\xi}{v_{F}(2\pi)^{3}}\left|U(\textbf{q})\langle
B|A\rangle_{\textbf{q}}\right|^{2}iD_{AB}(\varepsilon_{n}-\varepsilon_{m})i\widetilde{F}(\xi,\varepsilon_{m})\label{2.2a}\\
&&\widetilde{\varepsilon}_{n}=\varepsilon_{n}+\rho
T\sum_{A}\sum_{B}\varpi_{A}\sum_{m=-\infty}^{+\infty}\int\int\frac{d^{2}qd\xi}{v_{F}(2\pi)^{3}}\left|U(\textbf{q})\langle
B|A\rangle_{\textbf{q}}\right|^{2}iD_{AB}(\varepsilon_{n}-\varepsilon_{m})i\widetilde{G}(\xi,\varepsilon_{m}).\label{2.2b}
\end{eqnarray}
Eqs.(\ref{2.2a},\ref{2.2b}) are a set of self-consistent
equations. The order parameter $\Delta$ is determined by the
anomalous propagator $F$ in a case of a pure metal and determined
by the dressed anomalous propagator $\widetilde{F}$ in a case of a
duty metal:
\begin{eqnarray}
&&\Delta(\varepsilon_{n})=T\sum_{m=-\infty}^{+\infty}\int
\frac{d\textbf{q}}{(2\pi)^{3}}
iF(\varepsilon_{m},\textbf{p})|g(\textbf{p})|^{2}
iD_{\texttt{ph}}(\varepsilon_{m}-\varepsilon_{m},\textbf{p}-\textbf{q})\quad \texttt{(in a pure metal)}\label{2.2c}\\
&&\Delta(\varepsilon_{n})=T\sum_{m=-\infty}^{+\infty}\int
\frac{d\textbf{q}}{(2\pi)^{3}}
i\widetilde{F}(\varepsilon_{m},\textbf{p})|g(\textbf{p})|^{2}iD_{\texttt{ph}}(\varepsilon_{m}-\varepsilon_{m},\textbf{p}-\textbf{q})
\quad \texttt{(in a disordered metal)}\label{2.2d}.
\end{eqnarray}
Eqs.(\ref{2.2a},\ref{2.2b}) can be reduced to a following form
after integration over $\xi$:
\begin{eqnarray}
&&\widetilde{\Delta}_{n}=\Delta_{n}+
\sum_{m=-\infty}^{+\infty}W(n-m)\frac{\pi T\widetilde{\Delta}_{m}}{\sqrt{\widetilde{\varepsilon}_{m}^{2}+|\widetilde{\Delta}_{n}}|^{2}}\label{2.3a}\\
&&\widetilde{\varepsilon}_{n}=\varepsilon_{n}+\sum_{m=-\infty}^{+\infty}W(n-m)\frac{\pi
T\widetilde{\varepsilon}_{m}}{\sqrt{\widetilde{\varepsilon}_{m}^{2}+|\widetilde{\Delta}_{n}}|^{2}}\label{2.3b},
\end{eqnarray}
where
\begin{eqnarray}\label{2.4}
W(n-m)= \sum_{A}\sum_{B}\varpi_{A}\varpi_{B}\int\frac{2\rho
d^{2}q}{\omega_{AB}v_{F}(2\pi)^{3}}\left|U(\textbf{q})\langle
B|A\rangle_{\textbf{q}}\right|^{2}\frac{\omega_{AB}^{2}}{(n-m)^{2}\pi^{2}T^{2}+\omega_{AB}^{2}}
\end{eqnarray}
The gap $\widetilde{\Delta}_{m}$ is an even function of $2m+1$,
but the energy parameter $\widetilde{\varepsilon}_{m}$ is an odd
function of $2m+1$. Hence these function are renormalized in
different ways:
\begin{equation}\label{2.4a1}
\frac{\widetilde{\Delta}}{\widetilde{\varepsilon}}>\frac{\Delta}{\varepsilon}.
\end{equation}
From Eqs.(\ref{2.2c},\ref{2.2d}) and Eqs.(\ref{2.1}) we can see
that unequality (\ref{2.4a1}) ensures increasing of the gap
$\Delta$ as compared with a pure superconductor or with a dirty
superconductor with elastic impurities where an equality
$\frac{\widetilde{\Delta}}{\widetilde{\varepsilon}}=\frac{\Delta}{\varepsilon}$
takes place. Thus Anderson theorem is violated in the sense that
embedding of the impurities in $s$-wave superconductor increases
its critical temperature.

If temperature is much more than any impurity's frequencies
$T\gg\omega_{AB}$ then Eqs.(\ref{2.3a},\ref{2.3b}) have a form
\begin{eqnarray}
  &&\widetilde{\Delta}_{n}=\Delta_{n}+W(0)
\frac{\pi T\widetilde{\Delta}_{n}}{\sqrt{\widetilde{\varepsilon}_{n}^{2}+|\widetilde{\Delta}_{n}}|^{2}} \label{2.4a}\\
  &&\widetilde{\varepsilon}_{n}=\varepsilon_{n}+W(0)\frac{\pi
T\widetilde{\varepsilon}_{n}}{\sqrt{\widetilde{\varepsilon}_{n}^{2}+|\widetilde{\Delta}_{n}}|^{2}}
\label{2.4b}
\end{eqnarray}
Solving Eqs.(\ref{2.4a},\ref{2.4b}) we find that the gap and the
energy parameter are renormalized similarly:
\begin{equation}\label{2.5}
\frac{\widetilde{\Delta}}{\Delta}=\frac{\widetilde{\varepsilon_{n}}}{\varepsilon_{n}}=1+\frac{1}{2\tau}\frac{1}{\sqrt{\varepsilon_{n}^{2}+\Delta^{2}}}.
\end{equation}
The relation (\ref{2.5}) means realization of Anderson's theorem -
the gap and, accordingly, critical temperature do not change. The
limit $T\gg\omega_{AB}$ corresponds to an elastic scattering by
impurities with a scattering frequency $\frac{1}{2\pi\tau}$. It
should be noticed if the impurity's frequency is too large
$\omega_{AB}\rightarrow\infty$ then an interaction with the
impurities is weak $\sim 1/\omega_{AB}$ and effectiveness of the
impurities decreases.

Let us consider a case when temperature is equal to a critical
temperature $T=T_{\texttt{C}}^{\ast}$ of a system
metal+impurities. Then the gaps are equal to zero and
Eqs.(\ref{2.3a},\ref{2.3b}) have a form
\begin{eqnarray}
&&\widetilde{\Delta}_{n}=\Delta_{n}+
\sum_{m=-\infty}^{+\infty}W(n-m)\frac{\pi T\widetilde{\Delta}_{m}}{|\widetilde{\varepsilon}_{m}|}\label{2.6a}\\
&&\widetilde{\varepsilon}_{n}=\varepsilon_{n}+\sum_{m=-\infty}^{+\infty}W(n-m)\frac{\pi
T\widetilde{\varepsilon}_{m}}{|\widetilde{\varepsilon}_{m}|}\label{2.6b}.
\end{eqnarray}
Eq.(\ref{2.6b}) has an exact solution \cite{mahan}:
\begin{eqnarray}\label{2.7}
\widetilde{\varepsilon}_{n}=\eta_{n}\varepsilon_{n},\quad
\eta_{n}=1+\frac{1}{|2n+1|}\left[W(0)+2\sum_{l=1}^{|n|}W(l)\right].
\end{eqnarray}
To find a critical temperature of a pure superconductor
$T_{\texttt{C}}$ we have to solve Eliashberg equations when
$\Delta=0$ \cite{ginz,mahan}:
\begin{eqnarray}
&&Z(\varepsilon_{n})\Delta_{n}=\pi
T_{\texttt{C}}\sum_{|\varepsilon_{m}|\leq\omega_{c}}\left[L(n-m)-\mu^{\ast}\right]\frac{\Delta_{m}}{|\varepsilon_{m}|}\label{2.8a}\\
&&\left[1-Z(\varepsilon_{n})\right]\varepsilon_{n}=-\pi
T_{\texttt{C}}\sum_{m=-\infty}^{+\infty}L(n-m)\texttt{sign}\varepsilon_{m}\label{2.8b},
\end{eqnarray}
where $Z$ is a renormalization function,
\begin{eqnarray}\label{2.9}
L(n-m)=2\int^{\infty}_{0}d\Omega\alpha^{2}(\Omega)g(\Omega)\frac{\Omega}{\Omega^{2}+(\varepsilon_{n}-\varepsilon_{m})^{2}},
\end{eqnarray}
$\alpha^{2}(\Omega)g(\Omega)$ is an electron-phonon coupling
function, a restriction of summation over $m$  in (\ref{2.8a}) is
introduced to use Coulomb pseudopotential
\begin{eqnarray}\label{2.10}
\mu^{\ast}=\frac{\mu}{1+\mu\ln\left(\frac{E_{F}}{\omega_{c}}\right)}
\end{eqnarray}
instead of full Coulomb constant $\mu$, $\omega_{c}\sim
10\omega_{D}$ ($\omega_{D}$ is Debay frequency). Transition
temperature $T_{\texttt{C}}$ of the pure superconductor is a such
temperature when Eqs.(\ref{2.8a},\ref{2.8b}) have a solution.

To find a critical temperature of a system metal+impurities we
have to generalize Eliashberg equations. Electrons and Cooper
pairs scatter by impurities. As a result the gap and the energy
parameter are renormalized with Eqs.(\ref{2.6a},\ref{2.6b}).
$\Delta,\varepsilon\rightarrow\widetilde{\Delta},\widetilde{\varepsilon}$.
Then we have to substitute the renormalized function
$\widetilde{\Delta}_{m},\widetilde{\varepsilon}_{m}$ instead of
the functions $\Delta_{m},\varepsilon_{m}$ to the right side of
Eliashberg equations (\ref{2.8a},\ref{2.8b}). Then we have a set
of equations:
\begin{eqnarray}
&&Z_{n}\Delta_{n}=\sum_{|\varepsilon_{s}|\leq\omega_{c}}\left[L(n-s)-\mu^{\ast}\right]\frac{\widetilde{\Delta}_{s}}{|2s+1|\eta_{s}}\label{2.11a}\\
&&\widetilde{\Delta}_{n}=\Delta_{n}+
\sum_{m=-\infty}^{+\infty}W(n-m)\frac{\widetilde{\Delta}_{m}}{|2m+1|\eta_{m}}\label{2.11b}\\
&&Z_{n}=1+\frac{1}{|2n+1|}\left[L(0)+2\sum_{l=1}^{|n|}L(l)\right]\label{2.11c}\\
&&\eta_{n}=1+\frac{1}{|2n+1|}\left[W(0)+2\sum_{l=1}^{|n|}W(l)\right]\label{2.11d}
\end{eqnarray}
Two last formulas (\ref{2.11c},\ref{2.11d}) determine a
renormalization of electron specter due electron-phonon
interaction (the function $Z_{n}$) and due scattering by
impurities (the function $\eta_{n}$). Eq.(\ref{2.11b}) is a
nonhomogeneous set of linear equations in the unknowns
$\widetilde{\Delta}_{n}$ ($n=-\infty\ldots+\infty$) and the gap
$\widetilde{\Delta}$ is a function of the gap $\Delta$. After
subtetuting $\widetilde{\Delta}_{n}$ in Eq.(\ref{2.11a}) we have a
homogeneous set of linear equations in the unknowns $\Delta_{n}$.
Indexes of summation $n,s$ in Eq.(\ref{2.11a}) and $n,m$ in
Eq.(\ref{2.11b}) are independent. Temperature $T$ is contained in
the functions $L(n-s),W(n-m)$. Transition temperature
$T_{\texttt{C}}^{\ast}$ of the system is a such temperature when
Eqs.(\ref{2.11a},\ref{2.11b}) have a solution.

In order to consider an influence of impurities upon the
transition temperature we have to solve a homogeneous set of
equations obtained from Eq.(\ref{2.11b}) omitting $\Delta_{n}$:
\begin{eqnarray}\label{2.12}
\sum_{m}W(n-m)\frac{\widetilde{\Delta}_{m}}{|2m+1|\eta_{m}}-\widetilde{\Delta}_{n}=0
\end{eqnarray}
However Eq.(\ref{2.12}) has a solution at another temperature
$T^{\ast}$ - \textit{the singularity temperature} introduced in
\cite{grig}. The singularity temperature is
$T^{\ast}<T_{\texttt{C}}^{\ast}$ and it can be used as a lower
estimation of the critical temperature of the dirty metal. Its
physical sense is: the singularity temperature is a
superconducting transition temperature if we turn off the pairing
interaction in the metal. Therefore we have always
$T^{\ast}<T^{\ast}_{\texttt{C}}$. A determinant of the set of
equations (\ref{2.12}) must be equal to zero:
\begin{eqnarray}\label{2.13}
\texttt{det}D_{mn}(T_{\texttt{C}}^{\ast})=0,\quad
D_{mn}=\frac{W(n-m)}{|2m+1|\eta_{m}} -\delta_{mn},
\end{eqnarray}
where $\delta_{mn}=1$ if $m=n$, $\delta_{mn}=0$ if $m\neq n$.

Let an interaction with impurities is nonretarded (elastic):
$W(n-m)=W(0)\delta_{mn}$. The determinant $D_{mn}$ is diagonal in
this case. Each diagonal element of the determinant is
\begin{eqnarray}\label{2.14}
\frac{W(0)}{|2m+1|\eta_{m}}-1=\frac{W(0)}{|2m+1|+W(0)}-1\neq 0.
\end{eqnarray}
Hence the singularity temperature is absent. Indeed if the
interaction is elastic (when an addendum with $n=m$ is only) then
from Eqs.(\ref{2.11b},\ref{2.11d}) we can see
$\widetilde{\Delta}_{m}=\Delta_{m}\eta_{m}$. Then Eq.(\ref{2.11a})
is transformed to Eliashberg equation (\ref{2.8a}) for a pure
metal. Thus Anderson theorem is realized for the elastic
interaction: $T_{\texttt{C}}^{\ast}=T_{\texttt{C}}$. However if
the retarded interaction of quasiparticles with impurities takes
place then the transition temperature rises: $T^{\ast}\neq
0\Rightarrow T_{\texttt{C}}^{\ast}>T_{\texttt{C}}$ as a
consequence of dissimilar renormalizations of the gap and the
energy parameter:
$\frac{\widetilde{\Delta}}{\widetilde{\varepsilon}}>\frac{\Delta}{\varepsilon}$.
Namely the sign "$>$" of this inequality provides amplification of
the superconductive properties, unlike, for example, magnetic
impurities, where the sign is "$<$" resulting in suppression of
superconductivity.

\section{Conclusion.}\label{concl}

In this work a perturbation theory and a diagram technique has
been developed for a disordered metal if interaction of
quasiparticles with impurities is retarded and impurity's
oscillations are local. All possible diagrams are classified into
several types, the electron-impurity coupling for the various
impurity's transitions and mass operators for the basic scattering
processes are calculated: first order process (\ref{1.7}), second
order process (\ref{1.13},\ref{1.14}), higher-order cross-process
(\ref{1.16}), higher-order cross-process with involvement of
phonons (\ref{1.17}) and processes described with beam-like
diagrams (\ref{1.19}). We showed the perturbation theory can be
made in an adiabatic approximation (when impurity's transition
frequency is much less than metal's Fermi energy) for
cross-diagrams and in an approximation with uncoupled correlations
for the beam-like diagrams. Thus the electron-impurity coupling is
not assumed to be small unlike perturbation theory for the elastic
scattering. We found that in these approximations the averaging
over disorder results in a picture like quasiparticles interact
with some collective excitations propagating through the system,
thus an analogy between the inelastic scattering of electrons by
impurities and an electron-phonon interaction exists. In the
proposed diagram technique the lines of interaction with
impurities in the diagrams transfer both a momentum and an energy
parameter unlike the diagram technique for a disordered metal with
the elastic scattering. If the energy transfer cannot be then the
diagrams and their analytical representations are transformed into
diagrams and corresponding expressions for the elastic processes.
Thus the proposed perturbation theory generalizes a case of the
elastic scattering in a disordered metal.

Eliashberg equations at a critical temperature
$T_{\texttt{C}}^{\ast}$ have been generalized for a case of s-wave
superconductor containing impurities of a considered type:
Eqs.(\ref{2.11a}-\ref{2.11d}). We found the retarded interaction
of quasiparticles with impurities violates Anderson theorem: a gap
and an energy parameter are renormalizated differently
$\frac{\widetilde{\Delta}}{\widetilde{\varepsilon}}>\frac{\Delta}{\varepsilon}$
- Eq.(\ref{2.4a1}). This fact causes violation of Anderson's
theorem in the direction of increasing of the critical
temperature. Thus a critical temperature of a system
metal+impurity is more than a critical temperature of the pure
metal $T^{\ast}_{\texttt{C}}>T_{\texttt{C}}$. The increasing
depends on impurities' concentration, electron-impurity coupling
and oscillation specter of the impurities. Mechanism of influence
of an impurity on a Cooper pair is as follows: at first Cooper
pairs are formed in a metal with electron-phonon interaction, then
they are scattered by the impurities; the first electron changes
impurity's state, then the second one interacts with the impurity
changed by the first electron, thus a correlation between the
electrons appears that increases their binding energy. In a limit
case when temperature is much more then impurity's oscillation
frequency Anderson theorem is restored (effectiveness of the
impurities aspires to zero), because at too small frequency a
thermal noise destroys the changes of impurity's states. If the
frequency is too large ($\omega\sim\varepsilon_{F}\gg
T_{\texttt{C}}^{\ast}$) then an interaction with the impurities is
weak and effectiveness of the impurities decreases. The
generalized Eliashberg equation is simplified to Eq.(\ref{2.13})
if we calculate the singularity temperature $T^{\ast}$. Its
physical sense is: the singularity temperature is a
superconducting transition temperature if we turn off the pairing
interaction in the metal, therefore we have always
$T^{\ast}<T^{\ast}_{\texttt{C}}$. The singularity temperature we
can use as a lower estimate of the critical temperature of the
dirty superconductor.

\appendix

\section{Elastic scattering by impurities.} \label{A}

\begin{figure}[h]
\includegraphics[width=5cm]{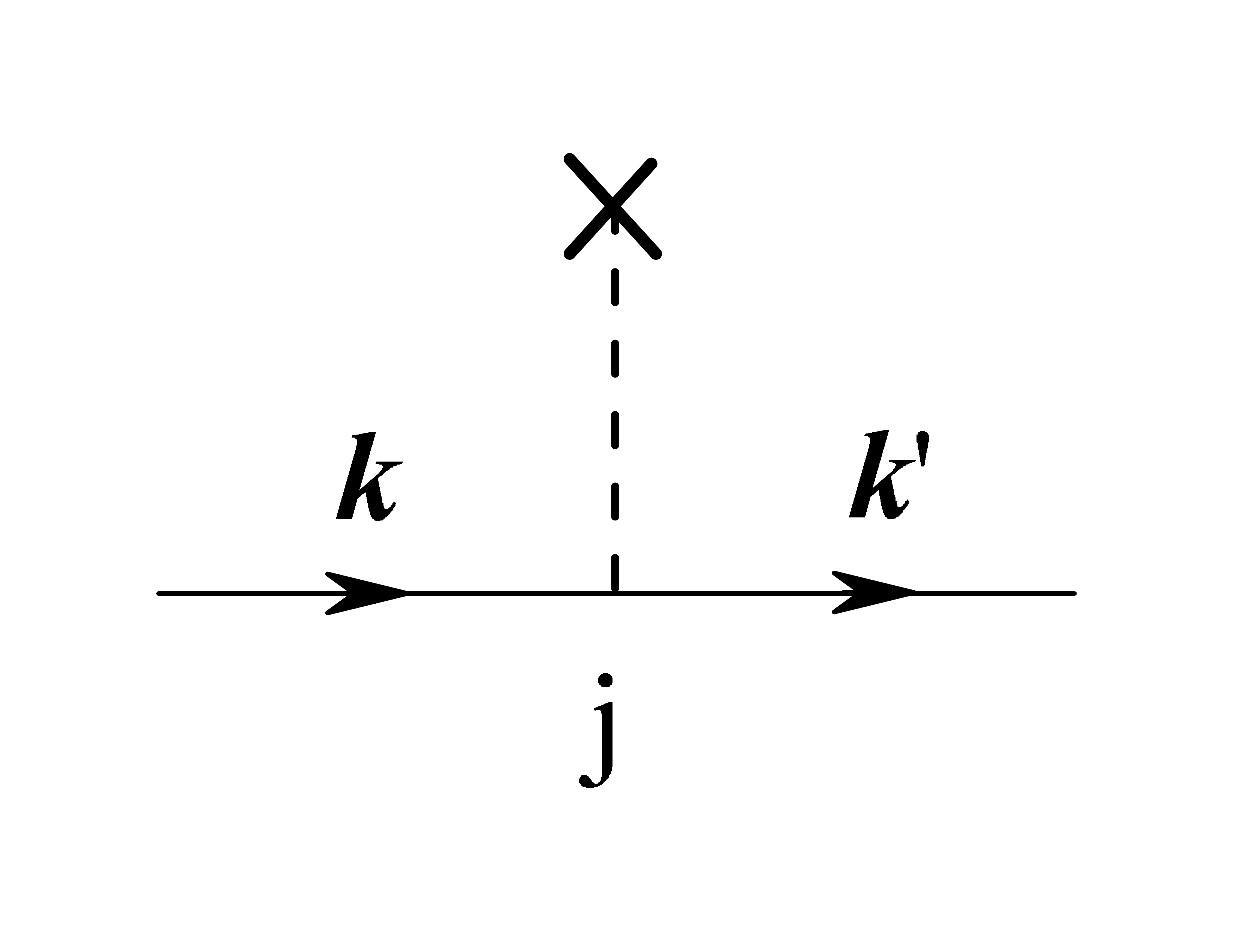}
\caption{The fist order diagram describing elastic scattering of a
quasiparticle by impurities.} \label{Fig1A}
\end{figure}

\begin{figure}[h]
\includegraphics[width=15cm]{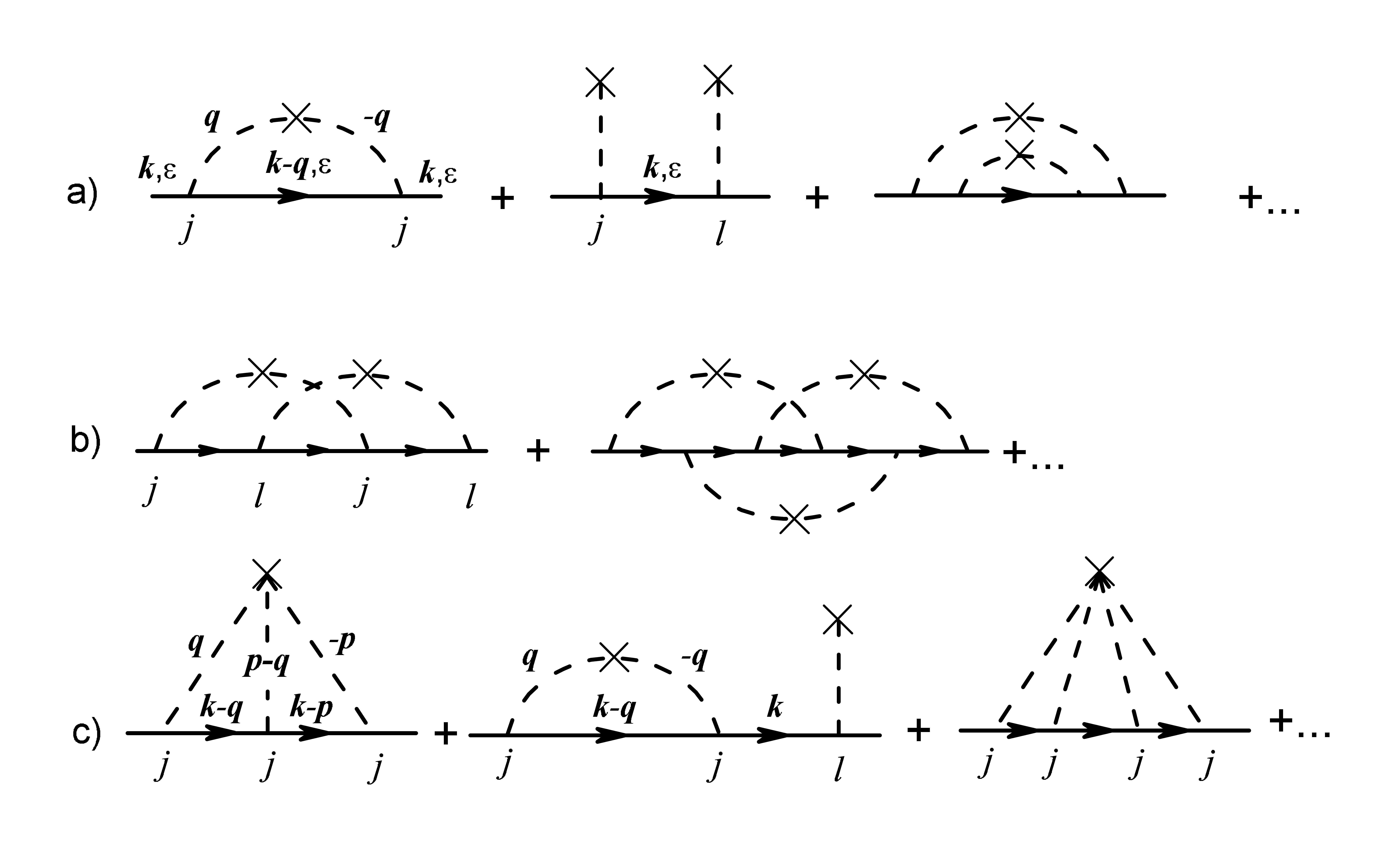}
\caption{Three kinds of diagrams describing an elastic scattering
of electrons by impurities. A kind (a) is usual diagrams
describing straight processes, a kind (b) is cross-diagrams and a
kind (c) is beam-like diagrams. $j$ and $l$ are deferent
impurities.} \label{Fig2A}
\end{figure}

In a case of elastic scattering of electrons by impurities
Hamiltonian of a system is
\begin{eqnarray}\label{A.1}
  \widehat{H} &=& \widehat{H}_{0}+\sum_{j}\sum_{\textbf{k},\textbf{k}'}\int\psi_{\textbf{k}'}^{+}(\textbf{r})
U(\textbf{r}-\textbf{R}_{j})\psi_{\textbf{k}}(\textbf{r})d\textbf{r}c^{+}_{\textbf{k}'}c_{\textbf{k}},
\end{eqnarray}
A summarized field of all impurities is
\begin{equation}\label{A.2}
    V(\textbf{r})=\sum_{j=1}^{N}U(\textbf{r}-\textbf{R}_{j})=
\frac{1}{V}\sum_{\textbf{q}}\sum_{j}U(\textbf{q})e^{i\textbf{q}\left(\textbf{r}-\textbf{R}_{j}\right)},
\end{equation}
the simplest process shown in Fig.\ref{Fig1A}.
The first correction to an electron propagator is 
\begin{eqnarray}\label{A.3}
&&iG_{1}(\textbf{k},\textbf{k}',\varepsilon)=\left[iG_{0}(\textbf{k},\varepsilon)\right]^{2}
(-i)\sum_{j}\int\psi_{\textbf{k}'}^{+}(\textbf{r})
U(\textbf{r}-\textbf{R}_{j})\psi_{\textbf{k}}(\textbf{r})d\textbf{r}\nonumber\\
&&=\left[iG_{0}(\textbf{k},\varepsilon)\right]^{2}
(-i)\frac{1}{V}\sum_{j}\int
e^{i(\textbf{k}-\textbf{k}')\textbf{r}}U(\textbf{r}-\textbf{R}_{j})d\textbf{r}
=\left[iG_{0}(\textbf{k},\varepsilon)\right]^{2}
(-i)\frac{N}{V^{2}}\int
e^{i(\textbf{k}-\textbf{k}')\textbf{r}}U(\textbf{r}-\textbf{R})d\textbf{r}d\textbf{R}\nonumber\\
&&=\left[iG_{0}(\textbf{k},\varepsilon)\right]^{2}
(-i)\frac{N}{V}(2\pi)^{3}\delta(\textbf{k}-\textbf{k}')\int
U(\textbf{r}-\textbf{R})d(\textbf{r}-\textbf{R})=
\left[iG_{0}(\textbf{k},\varepsilon)\right]^{2}
\rho(-i)U(q=0)(2\pi)^{3}\delta(\textbf{k}-\textbf{k}').
\end{eqnarray}
We can see the averaging operation leads to conservation of
momentum. In higher approximations we have diagrams shown in
Fig.\ref{Fig2A} (reducible and irreducible diagrams). There are
three kinds of the diagrams \cite{sad}. Types (a) and (b) are
diagrams corresponding to motion of an electron in Gauss random
field with factorized correlators (a white noise):
\begin{eqnarray}\label{A.4}
\left\langle V(\textbf{r}_{1})V(\textbf{r}_{2})\right\rangle=\rho
U^{2}\delta(\textbf{r}_{1}-\textbf{r}_{2}),\qquad \left\langle
V(1)\right\rangle=0,\qquad\left\langle
V(1)V(2)V(3)\right\rangle=0,\ldots\nonumber\\
\left\langle V(1)V(2)V(3)V(4)\right\rangle=\left\langle
V(1)V(2)\right\rangle\left\langle
V(3)V(4)\right\rangle+\left\langle
V(1)V(4)\right\rangle\left\langle V(2)V(3)\right\rangle+\ldots,
\end{eqnarray}
where in most cases an impurity's potential can be considered as
point so that $U(\textbf{q})\approx U=\int
U(\textbf{r})d\textbf{r}$. The type (b) corresponds to
cross-diagrams. A ratio of contribution of the cross-diagrams to
contribution of straight processes (the type (a)) is
$\frac{1}{lk_{F}}\ll 1$, where $l$ is a free length, $\ll 1$
corresponds to a week disorder. Diagrams of a type (b) we name
beam-like diagrams. In a limit
$\rho\rightarrow\infty,\upsilon^{2}\rightarrow
0,\rho\upsilon^{2}=\texttt{const}$ (continuous "spreading" of
impurities over a system) the beam-like diagrams disappear (except
reducible diagrams: for example the second diagram in a row (c)).

Conservation of momentum allows us to summarize diagrams with help
of Dyson equation:
\begin{equation}\label{A.5}
iG(\textbf{k},\varepsilon)=iG_{0}(\textbf{k},\varepsilon)+iG_{0}(\textbf{k},\varepsilon)(-i)\Sigma
iG(\textbf{k},\varepsilon),
\end{equation}
where  $\Sigma(\textbf{k},\varepsilon)$ is a mass operator. The
mass operator describes a multiple scattering of electrons by
impurities. For an elastic scattering the interaction lines do not
transfer energy parameter. They transfer momentum only. A
multiplier $\rho U(\textbf{q})^{2}$ is related to them. For a weak
disorder the mass operator is determined by the first diagram in a
row (a):
\begin{equation}\label{A.6}
(-i)\Sigma(\textbf{k},\varepsilon)=\rho\int\frac{d^{3}q}{(2\pi)^{3}}(-i)U(\textbf{q})
iG_{0}(\textbf{k}-\textbf{q},\varepsilon)(-i)U(-\textbf{q})=\rho\int\frac{d^{3}p}{(2\pi)^{3}}(-1)|U(\textbf{k}-\textbf{p})|^{2}
iG_{0}(\textbf{p},\varepsilon_{n}),
\end{equation}
or in Matsubara representation (nonzero temperature
$\varepsilon_{n}=(2n+1)\pi T$):
\begin{equation}\label{A.7}
-\Sigma(\textbf{k},\varepsilon_{n})=\rho\int\frac{d^{3}q}{(2\pi)^{3}}(-1)U(\textbf{q})
iG_{0}(\textbf{k}-\textbf{q},\varepsilon_{n})(-1)U(-\textbf{q})=\rho\int\frac{d^{3}p}{(2\pi)^{3}}|U(\textbf{k}-\textbf{p})|^{2}
iG_{0}(\textbf{p},\varepsilon_{n}).
\end{equation}
It should be noted that in the diagrams the dotted lines are not
dressed with polarization loops, because the disorder is "freezed
in" and the impurities do not fit into changes of an electron
density. Substituting a free propagator
$G_{0}(\textbf{p},\varepsilon_{n})=\frac{i}{i\varepsilon_{n}-\xi(p)}$
into the expression for a mass operator we obtaining (assuming a
weak dependence of a impurity's potential on momentum
$U(\textbf{k}-\textbf{p})\approx U$ and a linear specter of
quasi-particles near Fermi surface $\xi(k)\approx
v_{F}(k-k_{F})$):
\begin{eqnarray}\label{A.8}
\Sigma(\textbf{p},\varepsilon_{n})=-i\frac{\varepsilon_{n}}{|\varepsilon_{n}|}\pi\rho
U^{2}\nu_{F}\equiv-i\gamma \texttt{sign}\varepsilon_{n}
\end{eqnarray}
where $\nu_{F}=\frac{mk_{F}}{2\pi^{2}}$ is a density of states on
Fermi surface per one projection of spin. Then the mean free time
and the free length are determined as:
\begin{equation}\label{A.9}
    \tau=\frac{1}{2\gamma},\qquad
    l=v_{F}\tau=\frac{v_{F}}{2\gamma}=\frac{v_{F}}{2\pi\rho U^{2}\nu_{F}}
\end{equation}
Elastic impurities do not influence upon effective mass of
quasi-particles but they stipulate for a quasi-particles' damping
$\gamma\texttt{sign}\varepsilon_{n}$. It should be noticed
irreducible diagrams of kinds (a) and (c) can be summated in
$t$-matrix $t_{\textbf{k}\textbf{p}}$. Then in Eq.(\ref{A.7}) we
have to replace $\rho|U(\textbf{k}-\textbf{p})|^{2}$ by
$t_{\textbf{kp}}$ \cite{sad,loktev,pogor}.

All cross-diagrams describe quantum corrections for conductivity -
interference of incident and reflected by impurities electron
waves. This leads to Anderson's localization \cite{sad,sad1,pat}
when $\frac{1}{k_{F}l}\gtrsim 1$ - electrons are "blocked" between
the impurities. However with increase of temperature (or if the
system is in an external alternating field) nonelastic processes
begin to play a role (electron-phonon processes, electron-electron
processes) \cite{altsh1,altsh2,altsh3}. The processes limit the
coherence time of electron waves $\tau_{\varphi}<\infty$ (or the
coherence length $L_{\varphi}<\infty$). If $\tau_{\varphi}<\tau$
(or $L_{\varphi}<l$) then the interference contribution is
essentially suppressed because the phase failure takes place
\cite{levit}.

\end{document}